# Synthesis and structure of Na$^+$-intercalated WO$_3$(4,4′-bipyridyl)$_{0.5}$

Islah Udin[a,b], Matthew R. Fox[b], Hélène Martin[b,c]‡, Graeme J. Gainsford[b], John Kennedy[d], Andreas Markwitz[d], Shane G. Telfer, Geoffrey B. Jameson[a] and Jeffery L. Tallon[b]

**We have prepared single crystals of WO$_3$(4,4′–bipyridyl)$_{0.5}$ and doped these by Na-ion implantation. The structure of the resultant Na$_x$WO$_3$(4,4′–bipyridyl)$_{0.5}$ was determined by**
10 **single-crystal x-ray diffraction to comprise atomic layers of corner-shared WO$_5$N octahedra linked by the 4-4′-bypyridine via the apical nitrogen. In the observed space group of *Pbca*, the fully ordered bipyridyl molecules define cage-shaped structures, not the channels erroneously**
15 **reported previously for the *Cmca* polymorph. The Na ions are disordered bimodally about the cage centre, displaced in the *c*-direction so as to lie closer to the apical oxygens.**

Organic-inorganic hybrid materials offer a rich variety of structures with low-dimensional electronic properties and the
20 added benefit of synthesis at near-ambient conditions. One class of these materials is based on simple two-dimensional layers of tungsten (or molybdenum) oxide separated by organic molecules which either ionically or covalently bind the layers together[1]. The structure of tungsten oxide systems
25 is based on corner-and/or edge sharing WO$_6$ octahedra[2]. The insertion of group I or II cations into (or deoxygenation of) WO$_3$ and WO$_3$·nH$_2$O leads to an insulator-to-metal transition, produces the intense colour and metallic lustre of the tungsten bronzes, and induces bulk superconductivity in, for example,
30 Na$_x$WO$_3$ with $T_c$=3K for $x$=0.2[3]. Diffusion of Na into single crystals of WO$_3$ has even led to signatures of surface superconductivity around 91K[4]. The varied structural phases, electronic, physical and optical properties of these systems have been the subject of much research over past few
35 decades[5–8].

Recently, a related layered hybrid material WO$_3$(4,4′-bipyridyl)$_{0.5}$ was reported[9]. This consists of layers of single planes of corner-shared WO$_5$N octahedra with the axial organo-nitrogen (4,4′-bipyridyl) and oxo ligands of each
40 WO$_5$N octahedron alternating in direction, thereby binding the layers together. The bipyridyls were reported to align along the [110] direction thus defining open structural channels in this direction[9] and attracting our attention for ion implantation.

45
[a] *Institute of Fundamental Sciences, Massey University, Private Bag 11222 Palmerston North 4442, New Zealand.*
[b] *MacDiarmid Institute, Industrial Research Ltd., P.O. Box 31310, Lower*
50 *Hutt 5040, New Zealand. Tel: 64 4 9313 293; E-mail: j.tallon@irl.cri.nz*
[c] *École Nationale Supérieure de Chimie de Montpellier, 34296 Montpellier CEDEX 5, France.*
[d] *National Isotope Centre, GNS Science, P.O. Box 31312, Lower Hutt 5040, New Zealand.*
55 † Electronic Supplementary Information (ESI) available: For crystallographic files in .cif format see http://www.rsc.org/suppdata/cc/??/

In a preliminary study Na$^+$ was incorporated by aqueous electrochemistry[10] causing an increase in conductivity and the appearance of W$^{5+}$ species as evidenced by X-ray
60 Photoelectron Spectroscopy. In this paper we report the structural characterization of Na$_x$WO$_3$(4,4′-bipyridyl)$_{0.5}$ where the Na ions are intercalated by ion implantation. We find that in the parent compound ($x$=0) the bipyridyls define cages rather than the channels originally reported by Yan *et al*[9].
65 These studies confirm the bulk intercalation of Na$^+$ and open up the investigation of the *x*-dependent electronic phase behaviour of Na$_x$WO$_3$(4,4′-bipyridyl)$_{0.5}$ and related materials.

Bipyridyl tungstate was synthesized via hydrothermal reaction as described previously[9,10]. After cooling for five
70 hours to room temperature the hybrid material (in the form of fine yellow crystallites) was filtered and washed with ethanol and dried under vacuum. Crystals could even be prepared under ambient conditions but the crystallite size was smaller.

Sodium ions, $^{23}$Na$^+$, were implanted into both faces of
75 selected single crystals at 30 KeV using an lab-built low-energy ion implanter[11]. Crystal sizes were typically ~ (10-50)×(10-50)×(0.1-0.3) μm$^3$. Samples were implanted with Na$^+$ at estimated levels of 2% and 4% of the total number of atoms in the host material based on Monto Carlo calculations[12] using
80 the TRIM-DYNAMIC package. The calculation also indicated that most of the Na$^+$ ions resided in the top 120 nm of each surface and hence have penetrated the entire sample, though a bimodal concentration profile is likely. The pale yellow crystals darken to a green hue on ion implantation.

85 Single crystal X-ray diffraction (XRD) patterns were recorded with a Rigaku-Spider X-ray diffractometer with wrap-around image-plate detector and three-axis kappa goniostat. A number of single crystals were selected for size and quality, mounted on quartz fibres, characterised by single-
90 crystal XRD, then implanted while still mounted on the fibres and re-characterised by single-crystal XRD. Diffraction data for both un-implanted and implanted samples were collected at room temperature and low temperature. A problem with the cryocooling device prevented collection of the 4 % Na$^+$-
95 implanted sample at low temperature; in any event, earlier data collections had established the negligible changes in crystal properties on cooling. Thus for simplicity, detailed structural comparisons are made among the low-temperature structures (100 K) for unimplanted and 2 % implanted
100 samples (same crystal) and the room-temperature (20 ±0.2 °C) structure for the 4 % implanted sample (different crystal). Key crystal and refinement parameters are summarised in Table 1.

Unimplanted and Na$^+$-implanted crystals at 293 and 100 K are closely isomorphous. The asymmetric unit comprises a W
105 atom, two equatorial oxygen atoms, an axial oxygen atom and one ring of the bipyridyl group. Centres of inversion at the midpoint of the C4-C4' bond of the 4,4'-bipyridyl, 2$_1$ screw axes (parallel to *b*) and *c* glide planes generate the complete crystal structure, which comprises layers of corner-shared WO$_4$ groups in the *ab* [001] plane, separated by 4,4'-bipyridyl groups that coordinate axially to the W atoms and whose long axes are aligned approximately parallel to the *c* axis. Approximately octahedral coordination of the W atoms is completed by oxo groups. At room temperature (20 $^\circ$C) and relative to unimplanted crystals, Na$^+$ implantation led to a small increases in primarily the *b* and *c* axes and to a 0.8 % increase in unit cell volume. For the unimplanted crystal, no significant changes in unit-cell dimensions and volume (< 0.2%) were observed between 293 K and 100 K, but for the 2% Na$^+$-implanted crystal, there were decreases of 0.4 % and 0.6 % in the *a* and *c* axes respectively, leading to a decrease of 0.9 % in unit cell volume. Although no significant positional differences could be detected between room- and low-temperature structures for the unimplanted crystal, a significant decrease on cooling in atomic displacement parameters of the W atom was observed with $U_{equiv}$(W) decreasing from 0.0124(2) Å to 0.0087(2) Å; and other atoms show similar decreases, but much less significantly so. For the 2 % Na$^+$-implanted sample, a similar decrease was observed with $U_{equiv}$(W) decreasing from 0.0248(5) Å to 0.0146(3) Å. However, there was significant degradation in crystal quality following implantation, as indicated by increase in mosaicity of the 2 % Na$^+$-implanted crystal compared with the unimplanted crystal, as well as a small general increase in atomic displacement parameters ($U_{equiv}$(W) = 0.0146(3) Å, compared with 0.0124(2) Å).

**Table 1**. Selected refinement details WO$_3$(4,4′-bipyridyl)$_{0.5}$.

| Parameter | 0 % Na$^+$ | 2 % Na$^+$ | 4 % Na$^+$ |
|---|---|---|---|
| Temperature (/K) | 100 | 100 | 293 |
| $a$ (/Å) | 7.47709(13) | 7.4683(3) | 7.48501(14) |
| $b$ (/Å) | 7.38736(13) | 7.3910(3) | 7.40943(15) |
| $c$ (/Å) | 22.5625(4) | 22.5781(6) | 22.6257(4) |
| $V$ (/Å$^3$) | 1246.26(4) | 1246.27(8) | 1254.81(4) |
| Crystal system | Orthorhom. | Orthorhom. | Orthorhom. |
| Space group | *Pbca* | *Pbca* | *Pbca* |
| $R_{int}$, $R_{sigma}$ (/%) | 7.1, 5.6 | 9.4, 7.9 | 12.6, 9.7 |
| Asymm. unit | WO$_3$C$_5$H$_4$N(1) | (1)Na$_{0.125}$[a] | (1)Na$_{0.125}$[b] |
| $R$1(obs), #data $I>2\sigma(I)$ | 0.037, 1070 | 0.054, 1015 | 0.053, 1018 |
| $R$1(all), #data | 0.038, 1120 | 0.057, 1125 | 0.061, 1192 |
| $wR$2 (all data) | 0.091 | 0.128 | 0.131 |

[a] Refined $U = 0.18$ Å.  [b] Refined $U = 0.13$ Å

The axial W=O(oxo) and W-N(bipyridyl) bonds of WO$_5$N octahedra alternate in direction in both a and b directions, leading to a undulating WO$_4$ plane in the *ab* plane. The bipyridyl group and the axial oxygen atom are well ordered. The equatorial oxygen atoms are, however, stochastically disordered either side of the line joining pairs of W atoms. Residual electon density was concentrated near W atoms indicating possible unresolved disorder, concomitant with the

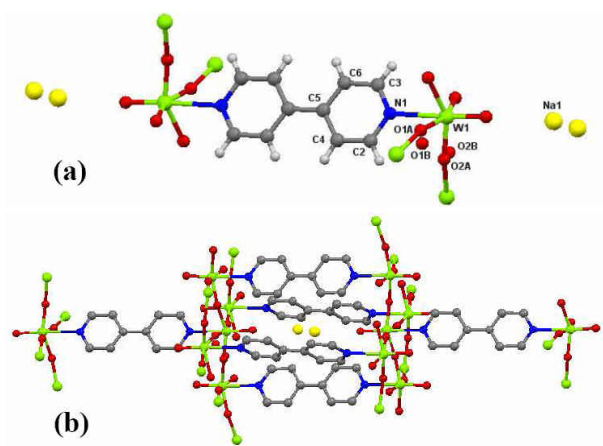

**Fig.1** (a) Structural repeat unit for Na$_x$WO$_3$(4,4′-bipyridyl)$_{0.5}$ showing two WO$_5$N octahedra linked via the nitrogens (blue) on the bipyridyl. The intercalated Na$^+$ ions (yellow) are bimodally located in the cage bounded along the *c*-axis by the axial oxygens (red), as depicted in (b). The cage structure is formed by the alternating alignment of the bipyridyl planes along [110] then [1$\bar{1}$0] directions. Tungsten atoms are shown in green. (*c*-axis is horizontal. Graphics using MERCURY[13]).

disorder of equatorial O atoms, or a missed superstructure. Close inspection of data frames, with calculated reflections displayed, revealed no missing weak reflections. As the residual electron density peaks were placed a distance from the W atom similar to the resolution of the data (0.81 Å), no attempt was made to model this disorder.

In both the 2 % and 4% Na$^+$-implanted samples, electron density, absent in the unimplanted structure, lying on the W=O line approximately 3.0 Å from the axial oxygen atom and disordered about a centre of inversion, was evident. This was assigned to the implanted Na$^+$ ion. The resulting structural unit is shown in Fig. 1. The consequences of ion implantation are mainly apparent in small changes in the slippage of one WO$_4$ layer relative to the next and in the displacement of the W atom from the plane of the pyridyl ring. The slippage of WO$_4$ layers, which also leads to a small and alternating tilt, layer-by-layer, of the bipyridyl group linking pairs of W atoms in the *c* direction, is greater parallel to the *b* axis (0.382 Å) than parallel to the *a* axis (0.162 Å). The slippage is perceptibly smaller for the Na$^+$-implanted samples: 2 % leading to 0.345 Å and 0.147 Å, respectively and 4 % leading to 0.358 Å and 0.082 Å. The effect of this slippage is to leave the W atoms slightly displaced from the plane of the pyridyl ring: for unimplanted, the displacement is 0.128 Å, for 2 % implanted 0.109 Å and for 4 % implanted 0.114 Å.

It is difficult to discern, because of disorder of equatorial oxygen positions, if implantation has caused any changes to the stereochemistry of the W octahedra. For the 2 % Na$^+$-implanted crystal there appears to be an increase in the asymmetry of bonds involving the W-O(1) bonds, which run approximately parallel to the *b* axis, whereas the significant asymmetry in the W-O(2) bonds, which run approximately parallel to the *a* axis, does not change. However, this change does not extend to the 4 % Na$^+$-implanted crystal.

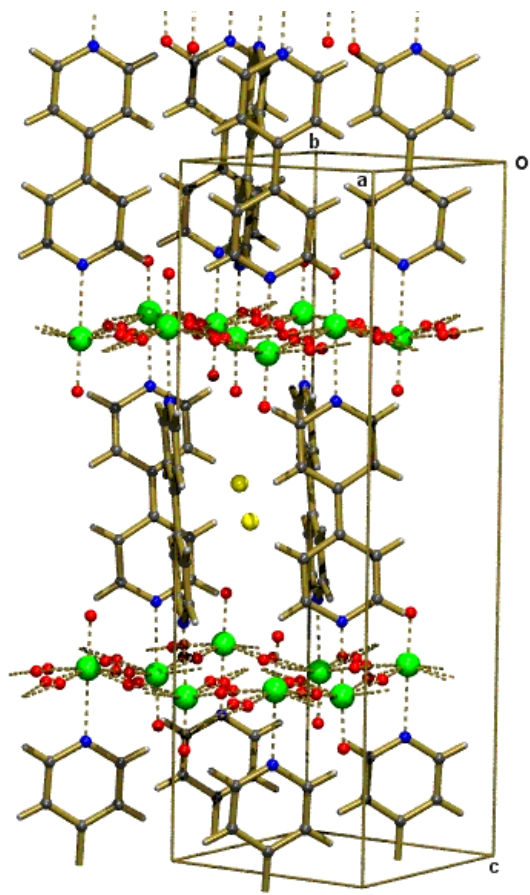

**Fig.2** The crystallographic structure of Na$_x$WO$_3$(4,4′-bipyridyl)$_{0.5}$ showing the inorganic WO$_3$N framework layers and one organic cage housing the Na$^+$ intercalant. Atoms are coloured as in Fig. 1. (Graphics using PLATON[14] and POV-RAY[15]).

The disposition of the Na$^+$ ions in the staggered cages formed by the bipyridyls is shown in Fig. 1(b). The orientational ordering of the bipyridyls with planes aligned alternately in the [110] then [1$\bar{1}$0] directions presumably arises from minimisation of proton repulsion between adjacent bipyridyls and from optimal space filling. The synthetic method used was that of Yan *et al.*[9], who reported the characterisation of their material in space group *Cmca*. We observe the same unit cell as Yan *et al.*[9], but observe instead systematic absences uniquely consistent with space group *Pbca*, an intriguing example of polymorphism that underscores the need for careful crystallographic analysis. Contrary to statements in their paper and to the published figure that there are channels parallel to the [110], we were disconcerted to discover, using the downloaded cif file (XEDGOC), that the bipyridyl rings, which lie parallel to (110), are in fact disordered by the crystallographic mirror plane perpendicular to the [100], and so as to fill space efficiently are likely to form cages similar to those that we observed. Similarly, in the description and diagram displaying the bipyrazine material, which notes channels parallel to [010], Yan *et al.*[9] fail to realise the disorder (and absence of channels) that is the ineluctable consequence of the tetragonal space group *I*4/*mmm*, which equivalences the [100] and [010] directions Interestingly, the bipyridyl material of Yan *et al.*[9] shows crystallographic equivalence in W-O-W bond lengths in the [010] direction and a marked asymmetry in the [100] direction, a pattern very similar to that which we observed for the unimplanted crystal. Curiously, whereas equatorial oxygen atoms are disordered and bipyridyl rings are ordered in space group *Pbca*, the converse is true in space group *Cmca*.

We have intercalated Na$^+$ ions into small single crystals of WO$_3$(4,4′-bipyridyl)$_{0.5}$ using a low-energy ion implanter and have determined the structure before and after implantation using single-crystal x-ray diffraction. The associated colour change reflects changes in the band structure, carrier doping and the appearance of W$^{5+}$ species[10]. The structure consists of planes of corner-shared WO$_5$N octahedra bound together by bipyridyl columns linking oxotungstate layers. The bipyridyls in the unimplanted WO$_3$(4,4′-bipyridyl)$_{0.5}$ are ordered along the [110] and [1$\bar{1}$0] directions so as to define cages rather than the [110] channels erroneously reported previously[9]. The intercalated Na$^+$ ions reside within these cages, distributed bimodally about the cage centre so as to lie closer to one or the other of the axial oxygens in the WO$_5$N octahedra. The generic M$_x$WO$_3$(4,4′-bipyridyl)$_{0.5}$ system (M=Na$^+$, K$^+$, Ca$^{++}$, etc.) and related materials represent a potentially phase-rich model system for exploring doping-dependent electronic properties and possible correlated states.

We acknowledge funding support from the Marsden Fund of New Zealand (grant IRL0501), the Allan Wilson Centre for Molecular Ecology and Evolution for the the rotating anode X-ray generator and the MacDiarmid Institute for Advanced Materials and Nanotechnology for the Spider X-ray detector and for research support. The Higher Education Comission, Pakistan, provided a doctoral scholarship for I.U.

## Notes and references

1   B. Ingham, S. V. Chong and J. L. Tallon, Organic-inorganic layered hybrid materials, in *Soft Condensed Matter: New Research*, ed. by K. I. Dillon, Nova Science Publishers (2007), p. 173-194.
2   P. J. Hagrman, R. L. LaDuca,, H. -J. Koo, R. Rarig, R. C. Haushalter, M. -H. Whangbo and J. Zubieta, *Inorg. Chem.*, 2000, **39**, 4311; B. Ingham, S. V. Chong and J. L. Tallon, J. Phys. Chem B 2005, **109**, 4936.
3   H. R. Shanks, Solid State Comm. 15 (1974), 753-756.
4   S. Reich and Y. Tsabba, *Eur. Phys. J.* B, 1999, **9**, 1.
5   M. Figlarz, *Prog. Solid State Chem*. 1989, **19**, 1.
6   A. K. Cheetham, C. N. R. Rao, and R. K. Feller, *Chem. Commun.*, 2006, 4780-4795.
7   B. Ingham, S. C. Hendy, S. V. Chong and J. L. Tallon, *Phys. Rev.* B, 2005, **72**, 075109.
8   G. Leftheriotis, S. Papaefthimiou, P. Yianoulis, *Solar Energy Mater. Solar Cells*, 2004, **83**, 115.
9   B. Yan, Y. Xu, N. K. Goh, L. S. Chia, *Chem. Commun.* (2000) 2169.
10  S. V. Chong, J. L. Tallon, *J. Phys. Chem. Solids* (submitted).
11  A. Markwitz and J. Kennedy, *Int. J. Nanotech*, 2009, **6**, 369.
12  J. P. Biersack, *Nuc. Instr. Meth. Phys. Res.* B, 1987, **19/20**, 32.
13  C. F. Macrae, P. R. Edgington, P. McCabe, E. Pidcock, G. Shields, R. Taylor, M. Towler and J. van de Streek, *J. Appl. Cryst.*, 2006, **39**, 453.
14  A. L. Spek, *J. Appl. Cryst.*, 2003, **36**, 7.
15  C. J. Cason, *POV-RAY* (2003). Persistence of Vision Raytracer Pty. Ltd, Victoria, Australia.